# Electronic excitations stabilised by a degenerate electron gas in semiconductors

C. Nenstiel,[1]* G. Callsen,[1]* F. Nippert,[1] T. Kure,[1] M. R. Wagner,[1] S. Schlichting,[1] N. Jankowski,[1] M. P. Hoffmann,[2] S. Fritze,[2] A. Dadgar,[2] A. Krost,[2] A. Hoffmann,[1] and F. Bechstedt[3]

[1]*Institut für Festkörperphysik, Technische Universität Berlin, Hardenbergstraße 36, 10623 Berlin, Germany*
[2]*Institut für Experimentelle Physik, Fakultät für Naturwissenschaften, Otto-von-Guericke-Universität Magdeburg, Universitätsplatz 2, 39016 Magdeburg, Germany*
[3]*Institut für Festkörpertheorie und -optik, Friedrich-Schiller-Universität, Max-Wien-Platz 1, 07743 Jena, Germany*

*\* nenstiel.christian@gmail.com, Gordon.Callsen@physik.tu-berlin.de*

Excitons in semiconductors and insulators consist of fermionic subsystems, electrons and holes, whose attractive interaction facilitates bound quasiparticles with quasi-bosonic character due to even-numbered pair spins. In the presence of a degenerate electron gas, such excitons dissociate due to free carrier screening, leaving a spectrally broad and faint optical signature behind. Contrary to this expected behaviour, we have discovered pronounced emission traces in bulk, germanium-doped GaN up to 100 K, mimicking excitonic behaviour at high free electron concentrations from $3.4\text{E}19/\text{cm}^3$ to $8.9\text{E}19/\text{cm}^3$. Consequently, we show that a degenerate, three-dimensional electron gas stabilizes a novel class of quasiparticles, named collexons, by many-particle effects dominated by exchange of electrons with the Fermi gas. The observation of collexons and their stabilisation with rising doping concentration, is facilitated by a superior crystal quality due to perfect substitution of the host atom with the dopant.

Many striking thermodynamic effects in nature are based on the reduction of the repulsive interaction between fermions or even on the bosonization of fermions and fermion complexes.[1] Prominent examples are macroscopic quantum phenomena such as superfluidity[2] and superconductivity.[3] Besides in the ground state, such phenomena can also be observed in excited states, e.g., of non-metallic matter with quasi-bosonic, excitonic excitations[4,5] in the low-excitation-density limit. Here, an ideal bosonic behaviour is prevented by electron-electron, hole-hole, and electron-hole interaction, hampering the system of interacting fermions to be condensed into a system of entirely massless, non-interacting bosons.[6,7]

Even more intriguing is the interaction of such quasi-bosonic excitons with a dedicated, degenerate Fermi system in its ground or even excited state - another fundamental phenomenon in many-body physics leading to exciton-cooling[8] or even the short-lived stabilisation ($\approx 25$ ps) of extended aggregates like the dropleton[9] in an electron-hole plasma at a temperature of 10 K. In contrast, excitons in bulk semiconductors commonly dissociate in the presence of a degenerate electron gas due to screening of the Coulomb interaction, leaving behind only the optical traces of band-to-band transitions.[10]

The influence of high free carrier concentrations on electronic states in heavily doped semiconductors is well-known.[10] Rising free carrier concentrations alter the optical characteristics of the semiconductor through the Burstein-Moss shift[11,12] (BMS), band gap renormalization[13] (BGR), and Pauli blocking of the optical transitions.[14]. In the limit of high free carrier concentrations, decreasing exciton binding energies facilitate a transition to Mahan-like excitons[15] at the Fermi edge singularity.[16] Excitonic excitations above the band edge have already been discussed for bulk systems[17] above the Mott density[18] - the onset of a metal-like state. Ultimately, the quasi-bosonic character of excitons is lost upon their dissociation into purely fermionic constituents (electron and hole). The spatial confinement inherent to *nanostructures*[19–21] like quantum dots[22,23] and quantum wells[8,24] stabilises quasiparticle complexes causing distinct optical features even under intermediate free carrier concentrations, with recent papers even reporting trions in two-dimensional crystals like $MoS_2$ layers.[25,26]

Here, we show that a novel class of exciton-like particles, named collexons, can increasingly be stabilised with rising density of a degenerate electron gas in a *bulk* semiconductor (GaN) up to 100 K due to many-particle effects. Upon optical excitation long-lived (> 250 ps), strongly localised quasiparticles are formed, which do not only comprise a single electron-hole pair, but also the collective excitation of the entire degenerate electron gas in the high-density limit. Contradicting the common behaviour of heavily doped semiconductors, the entire emission intensifies and its decay time slows towards increasing free electron concentration, while the collexonic emission narrows spectrally, as long no compensation occurs. Based on theoretical modelling of this intricate many-particle problem, we present a model that is consistent with all of our experimental findings, demonstrating an increasing bosonization of collexons with rising electron concentration. Our results confirm that, collexons are fundamental, many-particle excitations in the presence of a degenerate electron gas irrespective of the particular semiconductor host material.

## Photoluminescence signature of heavily doped GaN:Ge

**Figure 1** shows the PL signal of a highly germanium-doped (free electron concentration: n = 5.1E19/cm$^3$), n-type GaN film at a temperature of 1.8 K. Despite the present high doping level, two pronounced peaks stand out at $E_{N^*}^L$ = 3.492 eV and $E_N^U$ = 3.505 eV, labelled X$^L$ and X$^U$. We emphasize that emission lines with such low half width at half maximum (HWHM) values, with $\gamma_{N^*}^L$ = 4.50 ± 0.05 meV and $\gamma_N^U$ = 3.0 ± 0.1 meV, represent a highly unusual observation at the given doping level. We can exclude any luminescence contributions from the underlying buffer layer of pure GaN or inhomogeneities in the doped layer as discussed in the **Supplementary Information (SI)** under **Supplementary Note 1**.

A well-known characteristic band-to-band luminescence[27] evolves with the doping level as illustrated in the inset of **Fig. 1**. Here, the spectra range from a non-intentionally doped GaN (NID) reference sample (bottom, grey line) up to a free electron concentration of n = 8.9E19/cm$^3$ (top, green line) obtained by Hall effect measurements supported by an in-detail Raman analysis, cf. **Supplementary Note 2**. While the luminescence of the not-intentionally doped sample is still dominated by donor bound excitons,[28] any rise of the germanium concentration ($n_{Ge}$) by intentional Ge doping causes a spectral broadening as frequently observed for various semiconductor compounds.[29,30] Once the free electron concentration suffices, the corresponding Fermi energy passes the conduction band and causes above band gap luminescence due to the BMS, while the simultaneous extension of the luminescence towards lower energies is caused by the BGR, cf. **Fig. 1**. The corresponding critical carrier density is the Mott-density, at an electron concentration of about n = 7E18/cm$^3$ - 1E19/cm$^3$ in GaN.[27,31]

Several observations based on **Fig. 1** contradict the standard model for the behaviour of highly doped semiconductors. First the high doping level in the present case causes the appearance of distinct peaks (X$^L$ and X$^U$), and second the entire luminescence intensity in the band-edge region rises with $n_{Ge}$ due to an extraordinarily high crystal quality. Additionally, the lifetime $\tau_{BMS}$ of the BMS-shifted band-to-band transitions close to the Fermi edge increases with $n_{Ge}$ - a clear sign of negligible compensation (see **Supplementary Note 3**). A combination of SIMS, Hall-effect, Raman, transmission, reflection, and PL measurements (see **Supplementary Note 2**), proves that the present set of highly germanium-doped GaN samples exhibits only minor compensation effects as only a well-tolerable concentration of, e.g., extended structural defects and deep carrier traps is introduced.[32,33] Hence, the free electron concentration n scales with the dopant concentration $n_{Ge}$ in an almost linear fashion, with the highest doped sample as only exception, cf. **Supplementary Note 2**. The main reason for the overall high crystalline quality is given by the similarity of the $Ge^{4+}$ and $Ga^{3+}$ core radii and bond lengths in GaN.[34] The only significant doping-induced difference is the additional valence electron of Ge, that leads to the Fermi sea formation - a most ideal, textbook-like doping situation.

The effect of the high doping concentration in combination with low compensation is directly revealed by the particular luminescence characteristics of $X^L$ and $X^U$, cf. **Tab. I**. The spectral splitting between both peaks - the binding energy of $X^L$ ($E_{bind} = E_N^U - E_{N^*}^L$) in regard to $X^U$ - diminishes with rising free electron concentration. Simultaneously, the line-broadenings $\gamma_{N^*}^L(n)$ and $\gamma_N^U(n)$ (HWHM) decrease indicating a stabilisation mechanism. In addition, the time-resolved PL (TRPL) measurements from **Fig. 2a** reveal a characteristic scaling behaviour of the associated decay times along with pronounced quasiparticle transformation processes studied in more detail based on PL excitation (PLE) spectroscopy, cf. **Fig. 2b**. All these observations underline the novel characteristics of the elementary excitations that are not inhibited, but instead truly stabilised by the degenerate electron gas.

## Doping-driven balance of the fundamental excitations

The decay times of the $X^L$ and $X^U$ excitations surpass the value of the background BGR- and BMS-luminescence (compare **Supplementary Note 3**), indicating a localization mechanism induced by the Fermi sea of electrons. Interestingly, the decay times of the BGR- and BMS-luminescence increase with rising doping concentration (see **Supplementary Note 3**). Additionally the decay times $\tau_D^L$ and $\tau_D^U$ follow this trend, cf. **Tab. I**. While the high energy peak ($X^U$) features a fast rise time ($\tau_R^U < 10$ ps) in all transients shown in **Fig. 2a** (blue and red data points), its low energy counterpart ($X^L$) exhibits a well-resolvable rise time $\tau_R^L$ that diminishes with rising doping concentration, cf. **Tab. I**. The solid lines in **Fig. 2a** represent the results of a biexponential fitting model, taking into consideration rise and decay times (n = 3.4E19/cm$^3$ and 5.1E19/cm$^3$). While only the highest doped sample required the inclusion of a second decay time due to the onset of compensation and structural defect formation.[28]

**Figure 2b (top)** shows the PLE spectra with detection energies directly associated to $X^L$, $X^U$, and a defect-related feature in the regime of the BGR-luminescence unique to the sample with the highest free electron concentration (n = 8.9E19/cm$^3$). All differential PLE detection energies are assigned to the corresponding absolute PL energies by vertical, dashed lines, cf. **Fig. 2b (bottom)**. We show such differential PLE detection energies in order to eliminate the minor influence of varying strain levels in our samples as discussed in **Supplementary Note 2**. The PLE spectrum of $X^U$ (blue) features only one excitation channel comprising a hole from the A-valence band and an electron in the first excited state of the exciton-like complex - a situation similar to the excited A-exciton in high-quality, undoped GaN samples (see **Fig. S6**).[35–37] In contrast, the PLE spectra related to $X^L$ (red - colour gradient) appear as much less perturbed due to the enhanced localisation. That means, in the high energy regime, two additional excitation channels appear, comprising holes from the B- ($X_B^U$) and C- ($X_C^U$) valence band along with an electron in the ground state, cf. **Supplementary Note 2**. In addition, close-to resonant excitation of $X^U$ boosts the intensity of $X^L$, driving a quasiparticle transformation process, whose dynamics is quantified by $\tau_R^L$ extracted from the time-resolved analysis. Interestingly, the intensity of the excitation channel related to $X^U$ rises with doping concentration, while the entire PLE spectrum of $X^L$ gets more distinct. This observation, in addition to the particular scaling behaviour of the rise time ($\tau_R^L$) with doping concentration, supports the identification of $X^U$ and $X^L$ as the exciton-like collexon, and the four-particle complex, the bicollexon, both stabilised by the Fermi sea.

Shifting the PLE detection towards lower energies (e.g. 3.425 eV) yields a PLE spectrum that exhibits resonances in close energetic vicinity to $X^U$ and $X^L$. The energy shift in between the PLE resonances related to $X^U$ and $X^L$ and the corresponding luminescence features are clear indications for a light self-absorption process[28,38] as further described in **Supplementary Note 4**. Nevertheless, the assignment between the luminescence peaks and the corresponding excitation channels is apparent from **Fig. 2b**, showing that holes from the three topmost valence bands contribute to $X^L$, while the complexes related to $X^U$ possess an efficient conversion pathway towards $X^L$. Naturally, both complexes can dissociate and therefore contribute to any luminescence at lower energies as their components - electrons and holes - relax towards the corresponding defect bands, cf. **Fig. 2b**.

Trivial interpretations for $X^U$ and $X^L$ as, e.g., classical bound excitons related to germanium can be excluded. No discrete emission peaks with linewidths in the meV-regime should occur for this case because high dopant concentrations typically create a broad,[39] Ge-related defect band, causing a wide energy spectrum for bound excitons. Additionally, the reflectivity is heavily altered in energetic vicinity to $X^U$ and $X^L$, indicating strongly localised elementary excitations (see **Supplementary Note 4** for details). Their thermalisation behaviour even reveals the stability of $X^U$ and $X^L$ up to a temperature of 100 K (see **Supplementary Note 1 and 5**).

### Interpretation

All experimental observations correlate with the formation of many-particle complexes (see **Supplementary Methods 1** for details). The $X^U$ peak is identified as a complex interacting with the Fermi sea, achieving its stabilisation by exchange of electrons. Indeed, the most important energy gain mechanism providing stability in an electron gas is particle exchange.[14] This effect is illustrated in the upper, left panel of **Fig. 3a**. The second low energy peak $X^L$ is correspondingly interpreted as a more extended, many-particle complex, capturing an electron-hole pair from the Fermi sea (see upper, right panel of **Fig. 3a**). In contrast to the interband electron-hole pair forming the complex $X^U$, the collexon, the additionally captured electron-hole pair is of intraband nature. The resulting four-particle complex in interaction with the excited Fermi sea, the bicollexon, is consequently stabilized by electron exchange and Coulomb attraction. Its total excitation energy is lowered by the effective binding energy of the additional intraband electron-hole pair. Here, this effective value is the full binding energy minus the excitation energy of an electron from the Fermi sea above the Fermi level, leaving a hole behind.

In contrast to trions,[8,23–26] the bicollexon is electrically neutral, spinless, and of bosonic nature. PL measurements in presence of a magnetic field up to 5 T (not shown) do not reveal any shift or splitting of the collexon lines from **Figs. 1** and **2b**. We cannot exclude the formation of even larger complexes with more than one intraband electron-hole pair but only slightly smaller excitation energies in comparison to the bicollexon. Indeed, in **Fig. S9** the $X^L$ peak exhibits a prominent line-shape asymmetry below 10 K that may be traced back to additional, weakly bounded intraband electron-hole pairs. The theoretical picture developed for low or moderately doped semiconductor nanostructures as quantum wells, wires, or dots cannot be applied in the present case. The measured free electron concentrations (n = 3.4E19/cm$^3$, 5.1E19/cm$^3$, 8.9E19/cm$^3$) correspond to Fermi energies of $\varepsilon_F = \hbar^2 k_F^2 / 2m_e^* =$ 166, 217, and 315 meV in the conduction band with a Fermi wave vector of $k_F = (3\pi^2 n)^{1/3} =$ 1.00, 1.15, and 1.38 nm$^{-1}$, if an isotropic electron mass $m_e^* = 0.231\, m_e$ is assumed.[27] Direct comparison with the Bohr radius of an A-exciton in GaN ($a_{ex}$ = 2.16 nm) shows that the high-density limit $a_{ex} k_F \gg 1$ beyond the Mott density is reached, where free excitons and their charged, trionic counterparts are completely dissociated, nullifying their relevance for optical data in the high-density regime.

The extraordinary role of the degenerate electron gas for the formation of $X^U$ and $X^L$ is illustrated in **Fig. 3a**. The ground state (G) of the system is given by the Fermi sea in the $\Gamma_{7c}$ conduction band of GaN:Ge. Upon optical, inter-band excitation, the collexon is formed as an electron-hole pair well below the Burstein-Moss edge, but strongly *coupled* to the degenerate electron gas and *scattered* by the N electrons present in G state. The collexon forms a first excited state stabilized by the Fermi sea of electrons through unscreened electron exchange (illustrated by the arrows in **Fig. 3a**). It is protected against interactions with polar optical phonons, since the Fröhlich coupling[40,41] is also screened. The formation of the larger complex $X^L$ as a second excited state with a lower excitation energy is consequently explained by a two-step procedure. First, virtually, a collexon is excited. Second, scattering of the collexon in the Fermi sea excites an electron and leaves a hole behind, yielding a bicollexon complex, which emits at lower energies with respect to the collexon, cf. **Fig. 3a.** Finally, $X^L$ is a highly unconventional, mixed complex constituting intra- as well as interband excitations that originate from the optical excitation and the subsequent exciton-electron scattering ($\gamma_{U \to L}$) and coupling mechanism to the Fermi sea. The optical decay of a bicollexon leaves an intra-band excitation behind that will relax rapidly towards the G state - the unperturbed Fermi sea in the conduction band.

The particular formation process of the bicollexon $X^L$ explains the occurrence of the rise time ($\tau_R^L$) in the corresponding PL transients in **Fig. 2a** and the particular excitation channel in close energetic vicinity to the collexon $X^U$, cf. **Fig. 2b (top)**. The rising free electron concentration promotes the excitation channel associated to $X^U$ as the probability for electron exchange events is enhanced. The PLE spectra of $X^L$ (red - colour gradient) from **Fig. 2b (top)** confirm their correlation. The increasing stabilisation mechanism via electron exchange is quantified by $\tau_R^U$ and $\tau_R^L$ and their particular scaling behaviour with rising doping concentration (see **Tab. I)** in qualitative agreement with the spectral narrowing in **Fig. 1**.

**Modelling and discussion**

Even the most sophisticated theory focused on solving the Bethe-Salpeter equation is unsuitable to describe electron-hole excitation complexes with two or four particles coupled to the degenerate electron gas,[14,16] mainly due to the omission of vertex corrections to the integral kernel and non-particle conserving interactions (see **Supplementary Methods 1** for details). Therefore, an intuitive physically theory is developed. In the low excitation limit, the absorption and emission properties are dominated by bound states, e.g., of free 1s-excitons[14,42] or trions[43–45] and can be simulated by the imaginary part of a frequency-dependent optical susceptibility $\chi(\omega)$ as a sum of such quantities in oscillator form.[14,42,43] We follow this approach to describe the collexon and bicollexon by

$$\chi(\omega) = |M|^2 \left\{ N^U \frac{|\phi_N^U(0)|^2}{E_N^U(n) - \hbar\omega - i\gamma_N^U(n)} + N^L \frac{|\phi_{N^*}^L(0)|^2}{E_N^U(n) - \Sigma_{N^*}^L(n) - \hbar\omega - i\gamma_N^U(n)} \right\}. \quad (1)$$

Here, $E_N^U(n)$ and $\gamma_N^U(n)$ denote the density-dependent excitation energy and the line-broadening of the collexon ($X^U$), cf. **Tab. I**. The capture process of an additional electron-hole pair constituting the bicollexon ($X^L$) is described by a self-energy $\Sigma_{N^*}^L(n)$.[44,45] Its negative real part $E_{bind}(n) = -\text{Re}\,\Sigma_{N^*}^L(n)$ is the effective binding energy of the additional intraband electron-hole pair affected by the degenerate electron gas, $E_{N^*}^L(n) = E_N^U(n) + \text{Re}\,\Sigma_{N^*}^L(n)$, while the negative imaginary part $-\text{Im}\,\Sigma_{N^*}^L(n)$ characterises the modification of the bicollexon's line-broadening $\gamma_{N^*}^L(n) = \gamma_N^U(n) + \text{Im}\,\Sigma_{N^*}^L(n)$. The density-dependent envelope functions $\phi_N^U(0)$ and $\phi_{N^*}^L(0)$ of the many-particle complexes $X^U$ and $X^L$ appear in real space at the coordinate centre because the optical transition matrix element M is approximated by its value at the Γ-point.[14,42] Their squares characterise the localisation of the many-particle complexes with rising electron density. The index N labels the stabilizing degenerate gas with N electrons, the corresponding index $N^*$ indicates that the intraband electron gas is excited with an electron above the Fermi level and a hole in the Fermi sea, while the pre-factors $N^U$ and $N^L$ describe the number of corresponding complexes, cf. **Supplementary Methods 1**. Only the rate $\gamma_{U \to L}$, introduced in **Fig. 3a**, is directly accessible via TRPL measurements yielding $\tau_R^L$.

The modelling of PL spectra without the band-to-band contribution (compare **Fig. 1**) based on the fit model from **Eq. 1** is shown in **Fig. 3b**, considering slightly varying strain levels, cf. **Supplementary Note 2**. The most important parameters of the excitations are fitted self-consistently with the determined free electron concentration n. These are the relative positions of the two peaks $E_N^U$ and $E_{N^*}^L$, the line-broadenings $\gamma_N^U(n)$ and $\gamma_{N^*}^L(n)$, as well as the relative oscillator strengths $I_N^U(n) = N_U \left|\phi_N^U(0)\right|^2 \gamma_N^U(n)$ and $I_{N^*}^L(n) = N_L \left|\phi_{N^*}^L(0)\right|^2 \gamma_{N^*}^L(n)$ for the PL case. Despite the stabilization tendency due to the electron gas, $E_{bind}(n)$ of the additional intraband electron-hole pair decreases with rising n (see **Fig. 3b** and **Tab. I**) due to increasing intraband excitation energies that scale with the Fermi energy. The HWHM values of both emission lines, $\gamma_N^U(n)$ and $\gamma_{N^*}^L(n)$, reduce with rising n in accordance with a decay time prolongation - a statement particularly true for $X^U$ (see **Tab. I** and **Supplementary Note 2**). Naturally, the inverse of the decay times $\tau_D^{U/L}$ is always much smaller than $\gamma_N^{U/L}$ considering the time-energy uncertainty, due to parasitic, e.g., Auger and phononic process (see **Supplementary Methods 1**).

The entire set of lifetime and emission line widths trends unequivocally confirms the stabilizing action of the degenerate electron gas. The bicollexon emission line broadens in comparison to its collexon counterpart due to the increase of $Im \sum_{N^*}^L(n) = 1.0, 1.5, 1.9$ meV, cf. **Tab. I**. Interestingly, while the relative intensities $I_{N^*}^L(n)/I_N^U(n)$ remain almost constant with rising n in accordance with the PL spectra (see **Fig. 3b**), the bicollexon dominates the collexon peak despite of its higher complexity. This result directly illustrates the efficiency of the binding mechanism of the intra-band electron-hole pair, a process that can only be overcome under extreme optical pump conditions (> 10 MW/cm$^2$).

In summary, the fit of the PL spectra in **Fig. 3b** by means of **Eq. 1** confirms the identification of $X^U$ and $X^L$ as neutral and spinless electronic excitations, a collexon and a bicollexon, respectively, both stabilized by the degenerate electron gas.

After the doping-induced complete dissociation of the quasi-bosonic excitons into their fermionic constituents, a bosonic revival manifests due to the stabilising action of the degenerate electron gas. In general, the discovered two fundamental excitations should occur in the optical signature of any highly doped and at the same time lowly compensated semiconductor with an extraordinary high crystalline quality. No additional spatial confinement of carriers as in quantum well and quantum dot structures is needed.[44–46] In contrast, the formation of collexons is even favoured in bulk material due to the interaction with the 3D Fermi gas. Nevertheless, both, a sufficiently high doping level and low compensation do not suffice alone for the observation of collexons. Additionally the material quality of the undoped semiconductor must be high, which can be confirmed by the appearance of luminescence related to the free A- and B-exciton and low HWHM values for the Raman modes ($\approx 1.3 \pm 0.1$ cm$^{-1}$).[33] We believe that all, e.g., n-dopants that achieve a close-to-perfect substitution of the corresponding host atom should lead to near perfect, doped crystals comprising a degenerate electron gas. Other possible candidates are Ge dopants occupying Ga-sites in $Ga_2O_3$ or Sn dopants on In sites in InN and $In_2O_3$.

**Methods**

The GaN:Ge samples were grown by metalorganic vapour phase epitaxy (MOVPE) on (0001) sapphire substrates (0.25° off-oriented towards the m-direction) with germane as Ge source as reported by Fritze et al.[32] room temperature Hall effect measurements[33] were performed with standard van der Pauw method.

The photoluminescence (PL) and photoluminescence excitation (PLE) experiments were performed in one setup comprising a He-bath-cryostat providing a base-temperature of 1.8 K. For the continuous wave PL measurements, a HeCd-laser (325 nm) was used, while the PLE measurements were conducted with a dye-laser (100 Hz repetition rate, pumped by a XeCl excimer laser) containing 2-methyl-5-t-butyl-p-quaterphenyl (DMQ) as active medium. The luminescence signal was dispersed by an additive double monochromator (Spex 1404 - 0.85 m focal length, 1200 groves/mm, 500 nm blaze) equipped with an ultra-bialkali photomultiplier tube (Hamamatsu, H10720-210) . For the time-resolved PL (TRPL) measurements, the sample was excited with the fourth harmonic of a picosecond Nd:YAG laser (266 nm, 76 MHz repetition rate) in a Janis micro cryostat (ST-500) at a temperature of 5 K. The PL was spectrally and temporally analyzed by a subtractive double monochromator (McPherson 2035 - 35 cm focal length, 2400 groves /mm, 300 nm blaze) equipped with a multichannel-plate (MCP) photomultiplier (Hamamatsu R3809U-52) limiting the temporal resolution to $\approx 55$ ps. Standard photon counting electronics were applied in order to derive the final histograms. See **Supplementary Methods 2** for further experimental details.

Details regarding the theoretical treatment of electronic excitations in the Fermi sea can be found in the **Supplementary Methods**.

**Acknowledgements**

We acknowledge support from the Deutsche Forschungsgemeinschaft (DFG) within the Collaborative Research Center 787 (CRC 787). We thank, M. Müller und Prof. J. Christen for fruitful discussions and experimental support.

**Author contributions**

C.N. and G.C. contributed equally to this work, wrote the manuscript, and measured as well as interpreted the majority of all data. F.N. and T.K. provided significant experimental support and supported the scientific debate. S.S. and N.J. undertook the Cathodoluminescence measurements and analysis. M.P.H. recorded and analysed the SIMS data. S.F., A.D. and A.K. provided the GaN:Ge samples. F.B. performed the modelling and has a most significant contribution to the overall interpretation. A.H. strongly supported the interpretation of the experimental data and guided the project together with G.C., C.N., and F.B.

**References**

1. Gogolin, A. O., Nersesyan, A. A. & Tsvelik, A. M. *Bosonization and Strongly Correlated Systems*. (Cambridge University Press, 2004).
2. London, F. The λ-Phenomenon of Liquid Helium and the Bose-Einstein Degeneracy. *Nature* **141,** 643–644 (1938).
3. Bardeen, J., Cooper, L. N. & Schrieffer, J. R. Theory of superconductivity. *Phys. Rev.* **108,** 1175–1204 (1957).
4. Frenkel, J. On the transformation of light into heat in solids. *Phys. Rev.* **37,** 17–44 (1931).
5. Wannier, G. H. The structure of electronic excitation levels in insulating crystals. *Phys. Rev.* **52,** 191–197 (1937).
6. Blatt, J. M., Böer, K. W. & Brandt, W. Bose-einstein condensation of excitons. *Phys. Rev.* **126,** 1691–1692 (1962).
7. Kasprzak, J. *et al.* Bose-Einstein condensation of exciton polaritons. *Nature* **443,** 409–414 (2006).
8. Bigenwald, P., Kavokin, A. & Gil, B. Excitons and trions confined in quantum systems: From low to high injection regimes. *Phys. Status Solidi Appl. Res.* **195,** 587–591 (2003).
9. Almand-Hunter, A. E. *et al.* Quantum droplets of electrons and holes. *Nature* **506,** 471–475 (2014).
10. Shklovskii, B. I. & Efros, A. L. Electronic Properties of Doped Semiconductors. *Springer Ser. Solid-State Sci.* **45,** 72–93 (1984).
11. Burstein, E. Anomalous optical absorption limit in InSb. *Physical Review* **93,** 632–633 (1954).
12. Moss, T. S. The Interpreatation of the Properties of Indium Antimonide. *Proc. Phys. Soc. Sect. B* **67,** 775–782 (1954).
13. Berggren, K. F. & Sernelius, B. E. Band-gap narrowing in heavily doped many-valley semiconductors. *Phys. Rev. B* **24,** 1971–1986 (1981).
14. Bechstedt, F. *Many-Body Approach to Electronic Excitations*. (Springer, 2015).
15. Mahan, G. D. Excitons in degenerate semiconductors. *Phys. Rev.* **153,** 882–889 (1967).
16. Schleife, A., Rödl, C., Fuchs, F., Hannewald, K. & Bechstedt, F. Optical absorption in degenerately doped semiconductors: Mott transition or Mahan excitons? *Phys. Rev. Lett.* **107,** 236405 (2011).

17. Thomas, G. a. & Rice, T. M. Trions, molecules and excitons above the Mott density in Ge. *Solid State Commun.* **23,** 359–363 (1977).

18. Mott, N. F. Metal-insulator transition. *Rev. Mod. Phys.* **40,** 677–683 (1968).

19. Kheng, K. *et al.* Observation of negatively charged excitons X- in semiconductor quantum wells. *Phys. Rev. Lett.* **71,** 1752–1755 (1993).

20. Finkelstein, G., Shtrikman, H. & Bar-Joseph, I. Optical Spectroscopy of a Two-Dimensional Electron Gas near the Metal-Insulator Transition. *Physical Review Letters* **74,** 976–979 (1995).

21. Shields, A. J., Pepper, M., Ritchie, D. A., Simmons, M. Y. & Jones, G. A. C. Quenching of excitonic optical transitions by excess electrons in GaAs quantum wells. *Phys. Rev. B* **51,** 18049–18052 (1995).

22. Warming, T. *et al.* Hole-hole and electron-hole exchange interactions in single InAs/GaAs quantum dots. *Phys. Rev. B - Condens. Matter Mater. Phys.* **79,** 1–6 (2009).

23. Xu, X. Enhanced trion emission from colloidal quantum dots with photonic crystals by two-photon excitation. *Sci. Rep.* **3,** 3228 (2013).

24. Esser, A., Runge, E., Zimmermann, R. & Langbein, W. Photoluminescence and radiative lifetime of trions in GaAs quantum wells. *Phys. Rev. B - Condens. Matter Mater. Phys.* **62,** 8232–8239 (2000).

25. Lui, C. H. *et al.* Trion-induced negative photoconductivity in monolayer MoS2. *Phys. Rev. Lett.* **113,** 1–5 (2014).

26. Scheuschner, N. *et al.* Photoluminescence of freestanding single- and few-layer MoS 2. *Phys. Rev. B - Condens. Matter Mater. Phys.* **89,** 2–7 (2014).

27. Feneberg, M. *et al.* Band gap renormalization and Burstein-Moss effect in silicon- and germanium-doped wurtzite GaN up to 1020 cm−3. *Phys. Rev. B* **90,** 75203 (2014).

28. Callsen, G. *et al.* Optical signature of Mg-doped GaN: Transfer processes. *Phys. Rev. B* **86,** (2012).

29. Arnaudov, B., Paskova, T., Goldys, E., Evtimova, S. & Monemar, B. Modeling of the free-electron recombination band in emission spectra of highly conducting n-GaN. *Phys. Rev. B* **64,** 1–12 (2001).

30. Desheng, J., Makita, Y., Ploog, K. & Queisser, H. J. Electrical properties and photoluminescence of Te-doped GaAs grown by molecular beam epitaxy Electrical properties and photoluminescence of Te-doped GaAs grown by molecular beam epitaxy. *J. Appl. Phys.* **999,** 53 (2001).

31. Binet, F., Duboz, J. Y., Off, J. & Scholz, F. High-excitation photoluminescence in GaN: Hot-carrier effects and the Mott transition. *Phys. Rev. B* **60,** 4715 (1999).

32. Fritze, S. *et al.* High Si and Ge n-type doping of GaN doping - Limits and impact on stress. *Appl. Phys. Lett.* **100,** 122104 (2012).

33. Nenstiel, C. *et al.* Germanium - The superior dopant in n-type GaN. *Phys. Status Solidi - Rapid Res. Lett.* **9,** 716–721 (2015).

34. Bogusławski, P. & Bernholc, J. Doping properties of C, Si, and Ge impurities in GaN and AlN. *Phys. Rev. B* **56,** 9496–9505 (1997).

35. Volm, D. *et al.* Exciton fine structure in undoped GaN epitaxial films. *Phys. Rev. B* **53,** 16543–16550 (1996).

36. Rodina, A. *et al.* Free excitons in wurtzite GaN. *Phys. Rev. B* **64,** 11–14 (2001).

37. Alemu, A., Gil, B., Julier, M. & Nakamura, S. Optical properties of wurtzite GaN epilayers grown on A-plane sapphire. *Phys. Rev. B* **57,** 3761–3764 (1998).

38. Broser, I., Gutowski, J. & Riedel, R. Excitation spectroscopy of the donor-acceptor-pair luminescence in CdS. *Solid State Commun.* **49,** 445–449 (1984).

39. Taniyasu, Y., Kasu, M. & Makimoto, T. An aluminium nitride light-emitting diode with a wavelength of 210 nanometres. *Nature* **441,** 325–328 (2006).

40. Fröhlich, H., Pelzer, H. & Zienau, S. XX. Properties of slow electrons in polar materials. *Philos. Mag.* **41,** 221–242 (1950).

41. Fröhlich, H. Electrons in lattice fields. *Adv. Phys.* **3,** 325–361 (1954).

42. Haug, Hartmut (Goethe-Universität Frankfurt, Germany), Koch, Stephan W (Philipps-Universität Marburg, G. *Quantum Theory of the Optical and Electronic Properties of Semiconductors*. (World Scientific, 2009).

43. Esser, A., Zimmermann, R. & Runge, E. Theory of trion spectra in semiconductor nanostructures. *Phys. Status Solidi Basic Res.* **227,** 317–330 (2001).

44. Suris, R. A. *et al.* Excitons and trions modified by interaction with a two-dimensional electron gas. *Phys. Status Solidi Basic Res.* **227,** 343–352 (2001).

45. Suris, R. A. in *Optical Properties of 2D Systems with Interacting Electrons SE - 9* (eds. Ossau, W. & Suris, R.) **119,** 111–124 (Springer Netherlands, 2003).

46. Koudinov, A. V. *et al.* Suris tetrons: Possible spectroscopic evidence for four-particle optical excitations of a two-dimensional electron gas. *Phys. Rev. Lett.* **112,** 1–5 (2014).

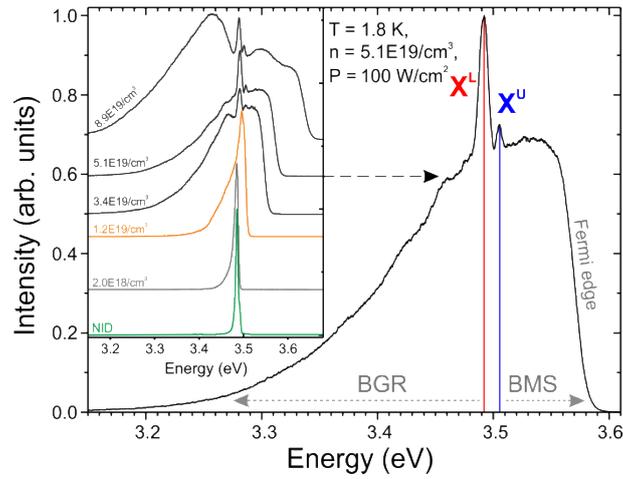

**Figure 1 | Exciton-like particle stabilization attested by photoluminescence.** Optical signature of a highly Ge-doped GaN sample showing the typical band gap renormalization (BGR) and Burstein-Moss-shift (BMS) along with two pronounced peaks $X^L$ and $X^U$ associated to many-particle complexes stabilized by the Fermi sea of electrons - the collexons. The inset shows a corresponding doping series starting with a non-intentionally doped (NID) sample featuring bound-excitonic emission. Upon rising doping concentration, the spectra broaden and the Fermi edge appears, before the two peaks $X^L$ and $X^U$ develop.

| Free carrier concentration n (× 1E19/cm$^3$) | $X^L$ position $E^L_{N^*}$ (eV) | $X^U$ position $E^U_N$ (eV) | $E^U_N - E^L_{N^*} =$ $E_{bind}$ (meV) ↓ | $\gamma^L_N(n)$ HWHM (meV) ↓ | $\gamma^U_N(n)$ HWHM (meV) ↓ | $\tau^L_D$ decay-time (ps) ↑ | $\tau^L_R$ rise-time (ps) ↓ | $\tau^U_D$ decay-time (ps) ↑ |
|---|---|---|---|---|---|---|---|---|
| 3.4 | 3.4914 | 3.5045 | 13.1 ± 0.3 | 4.4 ± 0.1 | 3.4 ± 0.2 | 303 ± 10 | 104 ± 10 | 267 ± 10 |
| 5.1 | 3.4923 | 3.5052 | 12.9 ± 0.3 | 4.5 ± 0.1 | 3.0 ± 0.1 | 344 ± 10 | 68 ± 10 | 281 ± 10 |
| 8.9 | 3.4897 | 3.5014 | 11.7 ± 0.3 | 3.9 ± 0.1 | 2.0 ± 0.1 | 303 ± 10 | < 10 | 305 ± 10 |

**Table I | Summary of the optical properties of $X^L$ and $X^U$.** The spectral positions of $X^L$ and $X^U$ ($E^L_{N^*}$ and $E^U_N$) extracted from Fig. 1 do not exhibit any particular scaling in regard to the free electron concentration due to varying strain levels in our layers, cf. Supplementary Note 3. Interestingly, the energetic splitting $E^U_N - E^L_{N^*} = E_{bind}$ diminishes with rising doping concentration along with the HWHM values $\gamma^L_N(n)$ and $\gamma^U_N(n)$ as indicated by the arrows. The decay-times $\tau^L_D$ and $\tau^U_D$ increase with rising doping concentration (except of $\tau^L_D$ at n = 8.9E19/cm$^3$, cf. Fig. 2a), while the rise-time $\tau^L_R$ associated to $X^L$ diminishes.

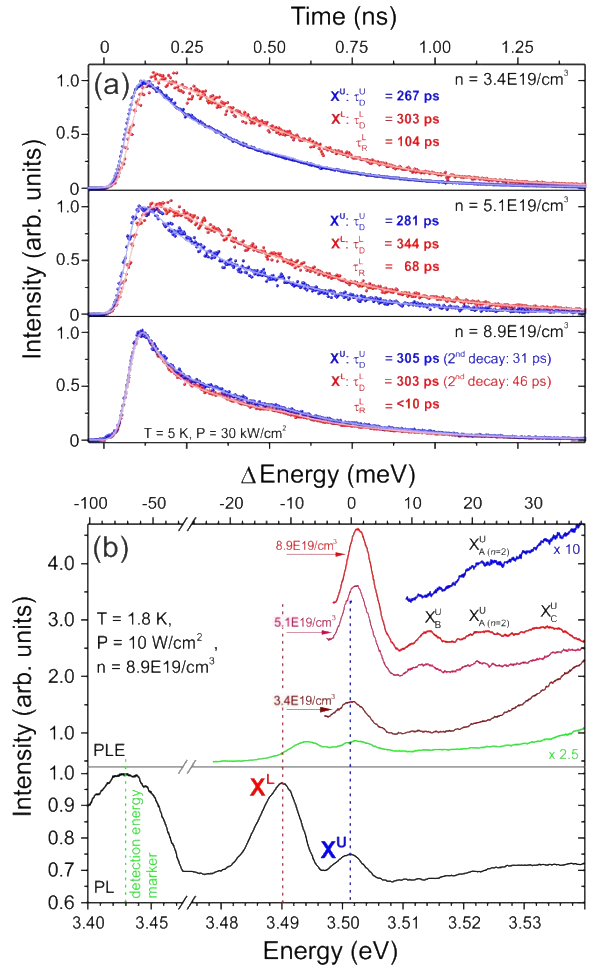

**Figure 2 | Doping-driven balance between the fundamental excitations $X^U$ and $X^L$.** (a) Time-resolved photoluminescence (PL) analysis of $X^L$ and $X^U$ for three doping concentrations. All transients (circles) can be approximated by a simple fitting model (solid lines) allowing to extract characteristic decay- ($\tau_D$) and rise-times ($\tau_R$). (b) Photoluminescence excitation (PLE) spectra (top) along with the corresponding photoluminescence spectrum (bottom) showing $X^L$ and $X^U$ along with the defect-related band at around 3.425 eV (n = 8.9E19/cm³). All PLE spectra show energy differences in regard to $E_N^U$ at the given free electron concentration in order to eliminate the influence of strain. While $X^U$ only shows a weak trace of a single excitation channel, $X^L$ features four excitation channels related to the topmost three valence bands of GaN and $X^U$. The emission band at 3.425 eV is excited predominantly via $X^L$ and $X^U$. Interestingly, an increase in the free electron concentration enhances the excitation channel related to $X^U \rightarrow X^L$ (the PLE spectra are not shifted).

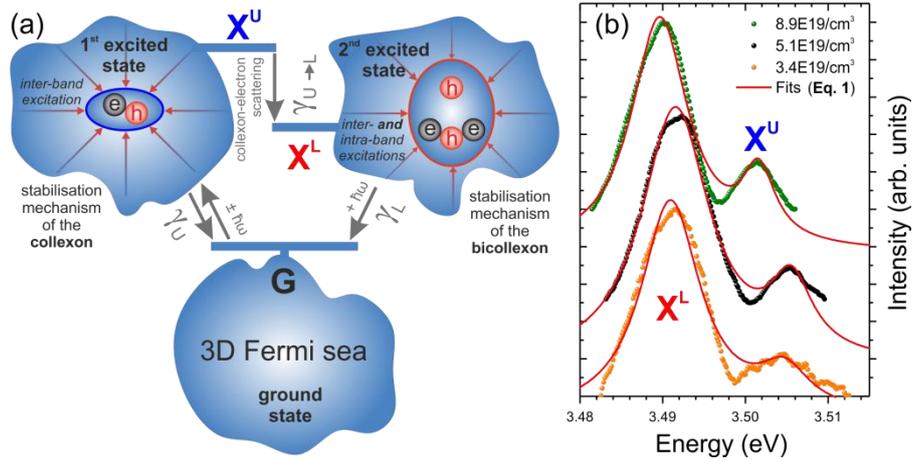

**Figure 3 | Collexonic excitations in the Fermi sea.** (a) The ground state (G) is formed by the degenerate electron gas - the Fermi sea - in the $\Gamma_{7c}$ conduction band of GaN:Ge. Upon optical excitation, inter-band excitations are formed and stabilized by the Fermi sea of electrons giving rise to the first excited state, the collexon ($X^U$) with a decay rate $\gamma_U$. In this high-density regime, such collexons scatter with electrons in the Fermi sea ($\gamma_{U\to L}$), enabling the formation of the bicollexon ($X^L$). Hence, the bi-collexon represents a mixed complex of inter- and intra-band electron-hole pairs that decays ($\gamma_L$), forming the second excited state. (b) Based on our modelling and the fitting function from **Eq. 1** (solid, red lines) we can approximate the Photoluminescence data (symbols) related to $X^U$ and $X^L$. All spectra are vertically displaced for clarity.

# Supplementary Information

**Electronic excitations stabilised by a degenerate electron gas in semiconductors**

C. Nenstiel,[1]* G. Callsen,[1]* F. Nippert,[1] T. Kure,[1] M. R. Wagner,[1] S. Schlichting,[1] N. Jankowski,[1] M. P. Hoffmann,[2] S. Fritze,[2] A. Dadgar,[2] A. Krost,[2] A. Hoffmann[1], and F. Bechstedt[3]

[1]*Institut für Festkörperphysik, Technische Universität Berlin, Hardenbergstraße 36, 10623 Berlin, Germany*
[2]*Institut für Experimentelle Physik, Fakultät für Naturwissenschaften, Otto-von-Guericke-Universität Magdeburg, Universitätsplatz 2, 39016 Magdeburg, Germany*
[3]*Institut für Festkörpertheorie und -optik, Friedrich-Schiller-Universität, Max-Wien-Platz 1, 07743 Jena, Germany*

*\* nenstiel.christian@gmail.com, Gordon.Callsen@physik.tu-berlin.de*

**SUPPLEMENTARY METHODS**

**1. Theoretical method**

The theoretical description of the novel many-particle complexes, collexon and bicollexon, stabilized by the presence of a degenerate electron gas from first principles requires the calculation of the spectral behaviour of the two- and four-particle Green functions, including the correct occupation of the quasiparticle bands. For the two-particle case this is possible by solving a corresponding Bethe-Salpeter equation (BSE).[1] Typically, Hedin's GW approximation[2] is applied, where for the description of quasi-electrons and quasi-holes vertex corrections to the single-particle self-energy are neglected, and only a statically screened Coulomb attraction between electrons and holes is taken into account.[3,4] In this limit, the BSE can also be solved in the presence of the degenerate electron gas.[5] This solution contains the band-gap renormalisation and the Burstein-Moss shift on the single-particle level[5–8] as well as the additional screening of the electron-hole interaction, which may lead to a dissociation of the exciton bound states and, hence, to the Mott transition[9], Fermi edge singularity, and Mahan excitons.[10] However, even this theory is unable to describe bound electron-hole excitations coupled to the degenerate electron gas with an excitation energy close to that of excitons in the absence of free carriers,[5] mainly due to the omission of vertex corrections in the BSE kernel.[1] The inclusion of the vertex function modified by the degenerate electron gas is a challenge for future theoretical investigations of the novel complexes. For more particle excitations such as in the case of biexcitons and trions in the low-density limit, the situation becomes worse as many-particle Green functions are required. Typically, such problems are approached constructing model Hamiltonians (see, e.g., Refs. 11,12). A T-matrix approximation containing particle-particle and particle-hole channels allows the description of charged excitons. Such approximations are also applicable in the biexciton case,[13] in particular to describe the additional binding between the two electron-hole pairs. Until now, there is no attempt to study the influence of a degenerate electron gas on such a four-particle excitation.

In order to describe collexon and bicollexon complexes additional approximations, based mainly on physical, but less on mathematical grounds, have to be introduced.

Since we are interested in near-band-edge optical phenomena of n-doped wurtzite GaN, we investigate only a two band model. Near the $\Gamma$-point only the $\Gamma_{7c}$ conduction band and the uppermost valence band $\Gamma_{9v}$ are taken into account. In the undoped case this restriction would be consistent with a study of the A-exciton. We assume the two bands to be isotropic as well as parabolic with the effective masses $m_e^*$ and $m_h^*$ and to be separated by a density-dependent direct gap $E_g(n)$.[6] Details regarding the band structure as the differences between the effective masses parallel and perpendicular to the c-axis[14] and the band coupling giving rise to the non-parabolicity[15] can be found elsewhere. The optical transition $\Gamma_{9v} \rightarrow \Gamma_{7c}$ is dipole-allowed at the $\Gamma$-point for light polarization perpendicular to the c-axis.[14] This polarization anisotropy will not be discussed further in the following discourse. Introducing the optical susceptibility $\chi(\omega)$ to describe PL and PLE spectra (see **Eq. 1** in the main article) we assume that the density-dependent envelope functions of the two complexes at vanishing particle distances are $\phi_{N^*}^L(0)$ and $\phi_N^U(0)$. This fact indicates that the sum of the oscillator strengths of the two complexes is not conserved for varying electron density as observed experimentally for trion and exciton.[16,17] In the collexon case it is described by the density-dependent Fourier-transformed envelope function, here given by:[3,12]

$$A_N^U(k) \sim \int d^3x \phi_N^U(x) e^{ikx}.$$

One may approximate this expression by the Fourier transform of a hydrogenic 1s-wave function

$$A_N^U(k) \sim \left[1 + \left(a_N^U(n)k\right)^2\right]^{-2}$$

with a characteristic radius $a_N^U(n)$. However, its density dependence strongly deviates from that of a screened Wannier-Mott exciton, which underlies a Mott transition in contrast to the collexon. Here, we have a complex whose characteristic extent is reduced by the exchange interaction with the electron gas. For the four-particle, bicollexon complex, three different relative motions appear. However, the most important one is the relative motion of the electron with respect to the corresponding hole in different bands. In order to illustrate the small extent of the complex, we similarly indicate this matter by assuming vanishing as indicated by $\phi_{N^*}^L(0)$, relative particle distances.

The localisation of the complexes in the presence of the electron gas leads to excitation energies $E_N^U(n)$ and $E_{N^*}^L(n)$ of the complexes, which are not coupled to the Burstein-Moss edge $E_g(n) + \varepsilon_F$ as in the high-density limit of Mahan excitons (compare **Fig. S1**). On the contrary, collexon and bicollexon appear below the excitation energy of an exciton in the absence of free carriers. This observation may be interpreted as a consequence of the unscreened particle-particle exchange with energies of the order of $\varepsilon_F$ for the electron-hole pairs excited virtually in the degenerate electron gas, thereby, almost compensating the Burstein-Moss shift.

Taking the collexon and bicollexon into account, the spectra in **Fig. 3b** can be approximated in the high density regime by

$$\text{Im } \chi(\omega) = |M|^2 \left\{ N^U \frac{|\phi_N^U(0)|^2 \gamma_N^U(n)}{[E_N^U(n) - \hbar\omega]^2 + [\gamma_N^U(n)]^2} + N^L \frac{|\phi_{N^*}^L(0)|^2 \gamma_{N^*}^L(n)}{[E_{N^*}^L(n) - \hbar\omega]^2 + [\gamma_{N^*}^L(n)]^2} \right\},$$

where collexon and bicollexon parameters, which are explained in the main article, are functions of the free carrier density (compare **Eq. 1** in the main manuscript).

The relation of the two many-particle complexes, collexon and bicollexon, can be discussed in two ways. The collexon may be considered as a dissociated bicollexon, where the intraband electron-hole pair recombines. We follow the more intuitive picture of a bound interband electron-hole pair stabilized by the electron gas, the collexon. Due to the strong interaction with the Fermi sea, the collexon may capture a low-energy electron-hole pair, whose binding in the complex overcomes its virtual excitation energy. This capturing is described by a pair self-energy $\sum_{N^*}^{L}(n)$ in **Eq. 1** (see the main manuscript). The self-energy $\sum_{N^*}^{L}(n)$ may be approximated in the framework of a T-matrix approximation with an effective attracting potential, similar to the treatment for quantum wells.[18,19]

Additionally, especially in GaN, it has to be taken into consideration that the emission linewidth of excitons is not trivially related to the inherent decay time. The particular spectral line shape depends on many factors such as, excitation energy, strain, and impurity density.[20] Specifically high doping densities lead to a pronounced broadening of spectral emission lines in bulk GaN as well as related nanostructures.[21,22] Starting at a free electron concentration of n = 3.4E19/cm$^3$, the emission linewidth of $X^U$ and $X^L$ diminishes as all detrimental effects of the host lattice - dictating the initial linewidths of $X^U$ and $X^L$ - are overcompensated by the beneficial stabilisation action of the Fermi sea of electrons. Hence, by increasing the free electron concentration, the weight for the origin of the observable linewidth is shifted from the crystal lattice to the degenerate electron gas. The rising doping concentration continuously weakens the quasi-bosonic character of all excitonic complexes until a re-bosonization is achieved in the form of $X^U$ and $X^L$ with rising free electron density and diminishing emission linewidths.

## 2. Supplementary experimental techniques

All (micro) Raman measurements were recorded with a Horiba LabRAM HR 800 (80 cm focal length, 1200 g/mm, 500 nm blaze) as described in Ref. 23 in detail. The Hall-effect measurements for determining the free electron concentration were undertaken in the standard Van-der-Pauw configuration. Secondary Ion Mass Spectrometry (SIMS) analysis was performed with an ATOMIKA 6500 quadrupole secondary ion mass spectrometer. Analysis for germanium was achieved using a Cs$^+$ primary beam with an incident energy of 14.5 keV and detection of negative secondary $^{74}$Ge ions. The 45 nA primary beam was typically scanned over a 200 μm x 200 μm area with ions detected from a region of 2500 μm$^2$ around the corresponding centre.

Low-temperature cathodoluminescence (CL) spectroscopy was directly performed in a modified JEOL 6400 scanning electron microscope (SEM) to study the optical properties of the individual GaN layers with a nanoscale spatial resolution.[24,25] For SEM-CL measurements, the electron beam was focused to a spot and either kept at a single position or scanned over the region of interest over the sample surface. Simultaneously, the emitted light is collected by an elliptical mirror and focused onto the entrance slit of a grating monochromator. Using Si-diode array intensified by a multi-channel plate, spectrally resolved CL imaging is performed. The measurements were operated at 5 keV and an electron-beam current of 110 pA

## SUPPLEMENTARY NOTES
### 1. Photoluminescence of the undoped GaN buffer

The luminescence signal introduced in **Fig. 1** is dominated by the highly Ge-doped GaN layer and does not show any contributions from the underlying GaN buffer layer. This statement is especially true for the two distinct luminescence signatures $X^U$ and $X^L$ in addition to the BGR and BMS luminescences that confirm the corresponding doping level. At the given excitation wavelength of 325 nm, the penetration depths for the laser light amounts to approx. 80 nm,[26] well below the GaN:Ge layer thickness of 700 nm (3.0E18/cm$^3$ - 5.1E19/cm$^3$). In order to further prove this point, the sample with the highest doping concentration (8.9E19/cm$^3$) exhibits a GaN:Ge layer thickness of 3.6 μm. Not even such large rise in GaN:Ge layer thickness affects, the appearance of $X^U$ and $X^L$ in the PL spectra (see the inset of **Fig. 1**), nor the particular, directly related scaling behaviours for several parameters as reported in **Tab. I** (decay- and rise-times, energetic positions, linewidths). In addition, a careful comparison of the luminescence related to the sole buffer layer (not intentionally doped - nid) and Ge-doped layers (e.g. 2.0E18/cm$^3$ and 5.1E19/cm$^3$) exhibits quite obvious differences as shown in **Fig. S1**. First of all, the not-intentionally doped (nid) GaN buffer layer's luminescence is dominated by bound-excitonic luminescence (DBX, oxygen-related[27]) in addition to free excitonic contributions (FX$_A$ and FX$_B$) at least one order of magnitude lower in intensity, cf. **Fig. S2**. As soon as the free electron concentration rises towards 2.0E18/cm$^3$, FX$_A$ and FX$_B$ fall under the detection limit, cf. **Fig. S1**. Any further rise in free electron concentration (e.g. towards 5.1E19/cm$^3$) re-establishes two peaks of comparable intensity (X$^U$ and X$^L$) in addition to BGR and BMS luminescence.

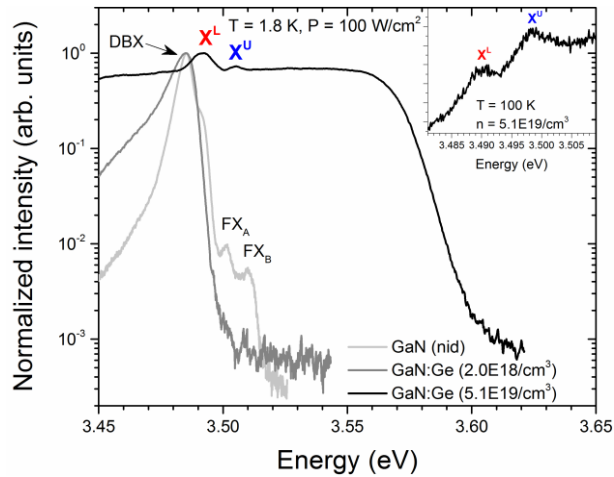

**Fig. S1 | Photoluminescence comparison.** The not-intentionally doped (nid) GaN layer exhibits a pronounced trace of donor-bound excitonic luminescence (DBX, oxygen-related) in addition to the signature of the free A and B exciton (FX$_A$ and FX$_B$) at least one order of magnitude lower in intensity, cf. **Fig. S2**. As soon as the electron concentration rises (2.0E18/cm$^3$), the free excitonic luminescence falls under the detection limit. Interestingly, any further increase in free electron concentration (5.1E19/cm$^3$) evokes the emission of X$^U$ at the energy position of FX$_A$ in the undoped GaN layer in addition to the appearance of X$^L$ in clear energetic separation to the DBX luminescence of the lower or not-intentionally doped samples. The inset shows that X$^L$ and X$^U$ are even still visible at a temperature of 100 K.

The most direct proof for the fact that X$^U$ and X$^L$ arise from the Ge-doped GaN layer appears from highly spatially resolved μPL and Cathodoluminescence (CL) measurements. Mapscans over the surface of the samples do not show any intensity inhomogeneities for X$^U$ and X$^L$ that could possible, e.g., be related to structural defects or clustering (not shown). In addition, highly spatially resolved cross-section CL spectra (see. **Fig. S2**) reveal that the highly Ge-doped layers are at least one order of magnitude brighter than the underlying GaN buffer layer, even if only the amplitudes are taken into account. For a better comparison **Fig. S2** includes the low excitation power PL spectrum (P = 1 W/mm$^2$) from **Fig. 1** associated to the highest doped sample (n = 8.9E19/cm$^3$), along with a counterpart recorded at elevated excitation powers (P = 100 W/mm$^2$). Here, the defect-related broad emission band at around 3.425 eV (compare **Fig. 2b (bottom)**) is strongly suppressed due to saturation effects, similar to the corresponding CL spectrum recorded perpendicular to the c-axis, cf. **Fig. S2**. Nevertheless, both emission peaks X$^L$ and X$^U$ are still present in the CL spectrum of the GaN:Ge layer in addition to the BGR and BMS luminescence. Naturally, the overall signal level in the cross-section CL measurements is lower in comparison to the PL measurements measured directly out of the c-plane of the GaN:Ge sample, as any cleavage of GaN introduces structural defects and relaxation. The latter point is directly evidenced by a global shift of $X^L$ and $X^U$ towards lower energies (< 5 meV) in the cross-section CL spectrum of the GaN:Ge layer shown in **Fig. S2**.

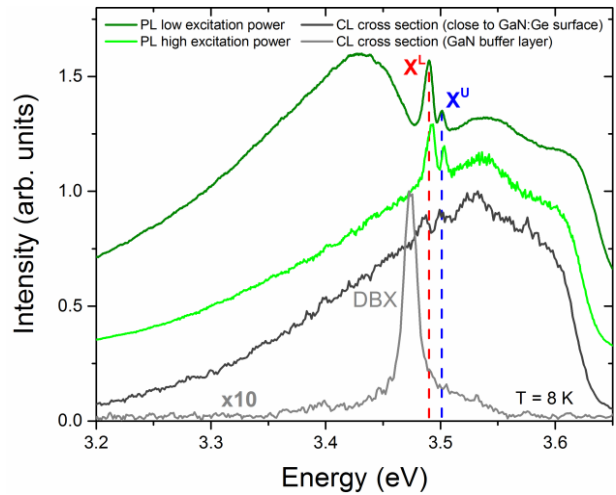

**Fig. S2 | Photoluminescence-Cathodoluminescence comparison.** The not-intentionally doped GaN buffer layer exhibits a cross-section (perpendicular to the c-axis) cathodoluminescence (CL) spectrum (grey) that is dominated by donor-bound exciton (DBX) emission, cf. **Fig. S1**. A comparable spectrum of the GaN:Ge layer close to the sample's surface shows a much stronger overall signal related to band-gap-renormalization, the Burstein-Moss shift, as well as $X^L$ and $X^U$. A direct comparison to a photoluminescence spectrum recorded at elevated excitation powers (light green, P = 100 W/mm$^2$) allows a straightforward association of all spectral features. Reducing the excitation power for the PL analysis (dark green, P = 1 W/mm$^2$) restores the emission characteristics shown in **Fig. 1**, along with a defect-related emission band at around 3.425 eV. Hence, none of the PL spectra reported in this manuscript for GaN:Ge are affect by any contributions from the GaN buffer layer underneath.

Depth-resolved CL measurements provide another useful method to further substantiate that $X^L$ and $X^U$ arise from the highly Ge-doped GaN layer. Here **Fig. S3** shows such CL spectra with varying excitation voltage and therefore electron penetration depths into the sample. The inset of **Fig. S3** illustrates the corresponding depth-dependency of the CL signal based on Monte-Carlo-Simulations. While at acceleration voltages up to ≈ 10 kV almost the entire CL signal arises from the GaN:Ge layers, any further increase in acceleration voltage shifts the origin of the CL signal towards the GaN buffer layer as shown in the inset of **Fig. S3**. Interestingly, despite the large change in acceleration voltage from 2 - 30 kV no change occurs for the CL signal. The luminescence of $X^U$ and $X^L$ is always observed in addition to the BGR and BMS luminescence as characteristics for the highly doped GaN:Ge layer. Hence, **Fig. S3** indirectly proves the main point from **Fig. S2** - the GaN buffer layer only exhibits a minor contribution to the overall CL signal that is always dominated by the GaN:Ge layer. As seen from **Fig. S3**, low CL acceleration voltages of, e.g., 2 kV approach the penetration depths of the UV laser light as discussed in **Supplementary Note 1**. Hence, the luminescence of $X^U$ and $X^L$ must always be related to the GaN:Ge layer, regardless of whether CL (**Fig. S2** and **S3**) or PL excitation (**Fig. 1**) is applied.

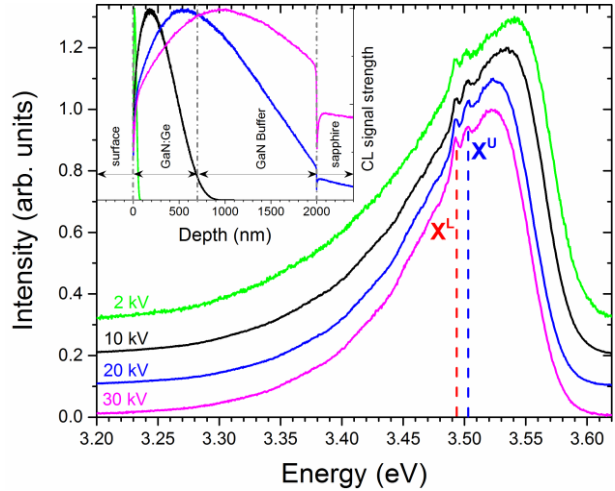

**Fig. S3 | Depth-resolved cathodoluminescence spectra.** Varying the acceleration voltage from 2 - 30 kV scans the origin of the cathodoluminescence (CL) signal through the sample's depth as shown in the inset based on Monte-Carlo-simulations. The luminescence related to $X^L$ and $X^U$ is always present at similar intensities independent on the electron penetration depth. Increasing the acceleration voltage beyond 10 kV might shift the potential origin of the entire CL signal towards the GaN buffer layer, however, no change in the overall CL signal is observed due to its weak luminescence contribution as already demonstrated based on cross-section CL spectra, cf. **Fig. S2**. Hence, $X^L$ and $X^U$ originate from the GaN:Ge layer.

Consequently, even for GaN:Ge layer thicknesses below 700 nm we would not expect any pronounced luminescence contributions of the GaN buffer for c-plane PL measurements similar to **Fig. 1**. Consequently, the particular luminescence signature of $X^U$ and $X^L$ arises from the GaN:Ge layer. Further experimental proof regarding the negligible influence of the GaN buffer layer in the PL spectra is given in **Supplementary Note 4** based on reflection data.

## 2. Compensation and strain in germanium-doped GaN

Our samples show no sign of any compensation mechanism up to a germanium concentration of $7E19/cm^3$ and only the onset of such an effect at $18E19/cm^3$. In order to illustrate this observation, **Fig. S4** correlates the germanium concentration of all our GaN:Ge samples determined by secondary ion mass spectrometry (SIMS) with the corresponding free electron concentration derived from three different experimental techniques, namely Photoluminescence and Raman spectroscopy (Fermi-edge shift[8] and LPP$^-$-mode[23] analysis) as well as Hall-effect measurements. These techniques were applied at different temperatures, but the increase of free carrier concentration over temperature is marginally in degenerated semiconductors.[23] As shown in **Fig. S4**, almost every Ge-atom contributes a free electron up to a germanium concentration of $7E19/cm^3$, before the onset of compensation and passivation occurs. Hence, our samples allow the study of an almost uncompensated semiconductor with free electron concentrations that approach the $1E20/cm^3$ regime. In order to substantiate this claim, the following **Supplementary Note 3** demonstrates that the Fermi-edge in the photoluminescence data experiences a constant shift towards higher energies with rising free carrier concentration - a phenomenon that is accompanied by rising decay-times, cf. **Supplementary Note 3**.

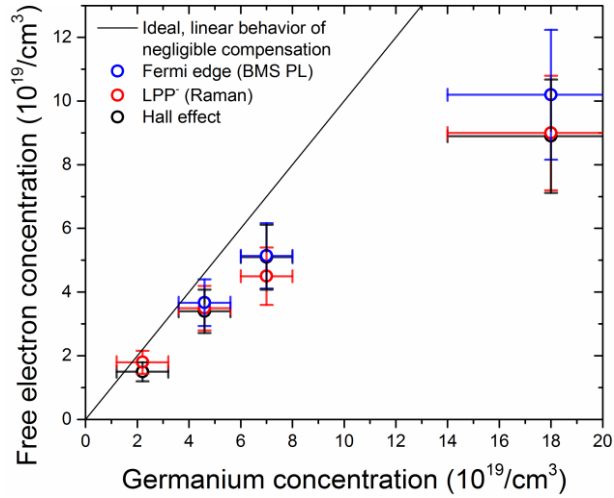

**Fig. S4 | Dopant vs. free electron concentration - compensation analysis.** Secondary ion mass spectrometry yield the germanium concentration in our samples. The corresponding free electron concentration is determined by three complementary techniques; namely Hall effect measurements, Raman (analysis of the longitudinal phonon plasmon – LPP⁻ coupling), and Photoluminescence (PL) spectroscopy (energetic position of the Fermi edge in the Burstein-Moss-shifted part of the luminescence). Interestingly, for germanium concentrations ($n_{Ge}$) of up to 7E19/cm³ we observe an almost linear rise with free electron concentration - a clear sign of negligible compensation. Only the sample with the highest free carrier concentration n = 8.9E19/cm³ starts to show compensation in direct agreement with the time-resolved data from **Fig. 2a**.

Further evidence for the low compensation level in our GaN:Ge samples is provided by photoluminescence and transmission measurements over a broad energy interval as shown in **Fig. S5**. For instance, the sample with the second highest germanium concentration with a free electron concentration of 5.1E19/cm³ exhibits an almost identical transmission up to the Fermi-edge if compared to the GaN:nid reference sample with the same thickness. Please note that the transmission modulation in **Fig. S5** is caused by the particular layer stack of our samples. No trace of any defect-induced absorption in the visible range is noticeable[28] for this free electron concentration (5.1E19/cm³) as the transmission features an almost flat characteristic over the broad energy interval from **Fig. S5**. A multitude of doping approaches is known to cause the, so-called, yellow and blue defect luminescence in GaN,[29] which are both not pronounced in our sample, cf. **Fig. S5**. The GaN:Ge sample with the highest free electron concentration (8.9E19/cm³) exhibits a ≈ 15 % reduced transmission (not shown) due to the onset of compensation and defect formation (e.g. induced by V-pits).

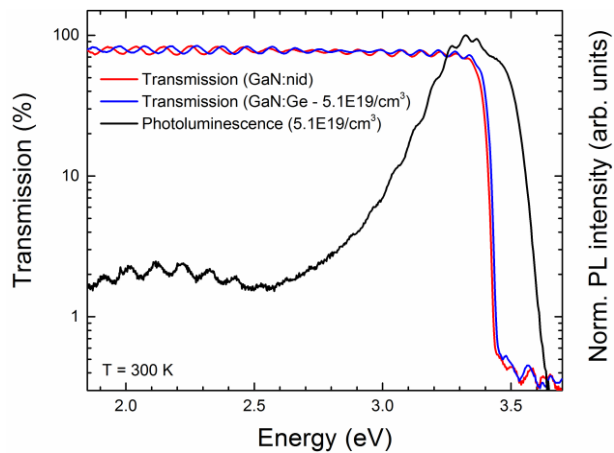

**Fig. S5 | Transmission and Photoluminescence characteristics.** The sample with a doping concentration of 5.1E19/cm³ is close to transparent over a broad energy range, only limited by the Fermi-edge at a temperature of 300 K. No absorption related to yellow or blue luminescence can be observed as most directly proved by the corresponding photoluminescence data recorded at a temperature of 5 K.

Nevertheless, the entire transmission characteristic is still comparably flat over a wide energy interval without any distinct traces of defect-related bands that would lower the transmission. Hence, the level of compensation due to the influence of defects is minor in our samples. Only the sample with the highest doping concentration marks the onset of compensation (see **Fig. S4**) as further discussed in the context of **Fig. 2** and in **Supplementary Note 3**, while an extraordinary high level of transmission is maintained.

The strain variation in our samples with germanium concentration is discussed in Ref. 23 in detail based on Raman spectroscopy. Here, the $E_2^{high}$ mode can be used to derive, the corresponding strain tensor component $\varepsilon_{zz}$ based on a set of stiffness constants[30] and phonon deformation potentials.[31] **Fig. S6** correlates $\varepsilon_{zz}$ for our non-intentionally doped (nid) GaN sample (triangles) with the free exciton transition energies related to the A- (grey), B- (light red), and C- (light blue) valence band (VB) in addition to the first excited-state of the A-exciton (light green, n = 2). For the GaN:nid sample we observe perfect agreement between the free exciton transition energies derived from PL and PLE measurements. In contrast, all Ge-doped samples feature an energetic offset in between the PL- and PLE resonances as described in more detail in **Supplementary Note 4**.

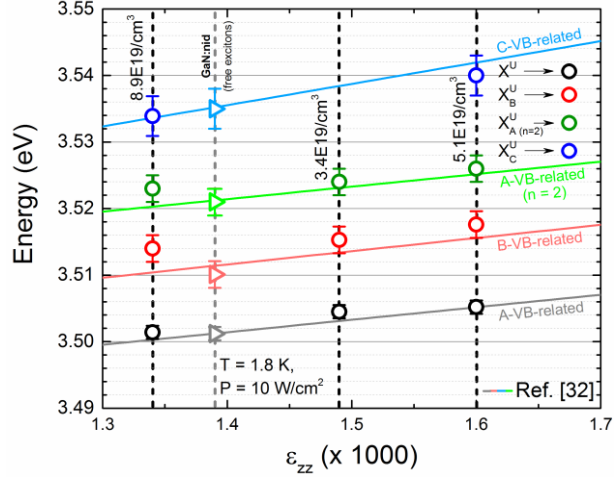

**Fig. S6 | Strain analysis.** Based on Raman spectroscopy we quantify the strain state of our samples expressed by $\varepsilon_{zz}$ and can consequently derive the energetic position of luminescence features related to the A- (grey), B- (light red), and C- (light blue) valence band (VB) based on Ref. 32 (solid lines). The directly related energetic position of free excitonic luminescence of our not-intentionally doped (nid) GaN sample is shown (triangles) in addition to its counterpart originating from the highly Ge-doped samples (circles). It is apparent, that the energetic position of all exciton-like resonances is dominated by the strain state in our samples. Please see the text for further details.

Here, **Fig. S6** summarizes the energy of all $X^U$-related excitation channels from PLE measurements (circles) we observe for our GaN:Ge samples, in direct relation to the strain tensor element $\varepsilon_{zz}$. As introduced in **Fig. 2**, these excitation channels of $X^U$ stand in close relation to the three topmost valence bands of GaN and are labelled accordingly with $X_B^U$, $X_{n=2}^U$ and $X_C^U$. Here, $X_{n=2}^U$ again denotes the excited state excitation channel (n = 2), related to the A-valence band. Based on Ref. 32, all $\varepsilon_{zz}$ values can be translated into corresponding free-exciton emission energies as shown in **Fig. S6** (solid lines). The Ge doped samples with free carrier concentrations below 3.4E19/cm³ do not show any similar luminescence features. As a result of this comparison in **Fig. S6**, it is apparent that the absolute energetic positions of all PLE excitation channels are dominated by the varying strain level in our samples - an observation that also directly applies for all emission energies derived from PL measurements. Nevertheless, especially for the sample with the highest free electron concentration (8.9E19/cm³) one observes an additional shift of the excitation channels towards higher energies in comparison to the expectations based on the strain level determined by Raman spectroscopy (solid lines). We explain this phenomenon that is directly related to the offset between PL and PLE resonances **in Supplementary Note 4**. In addition any offsets in **Fig. S6** are caused by the fundamentally different physical origin of free excitons and collexons leading to deviating, electronic deformation potentials. However, for both cases we expect a most prominent influence for the strain dependence of the valence bands that is directly demonstrated by the comparison shown in **Fig. S6**.

The effect of the varying strain level in our samples is partially reflected by the linewidths trend related to $X^L$ (see **Tab. I**) that is influenced by the strain level in the GaN:Ge layers and the free electron concentration. Within the present, minor strain variation[23] the energetic half-width of $X^L$ rises with increasing biaxial stress. Wagner et al. found a similar behaviour for bound and free excitons in ZnO.[33,34] Therefore, the linewidths of $X^L$ exhibits a maximum for the sample with a free electron concentration of $5.1E19/cm^3$ as a maximal bisotropic, bi-axial compressive stress of 1.19 GPa occurs. Any further rise of the free electron concentration leads to a decrease in the linewidths as the beneficial stabilisation effect by exchange overcompensates the detrimental strain effects.

**3. Time-resolved Photoluminescence analysis**

The transients for $X^L$ and $X^U$ from **Fig. 2a** are differential transients that are not affected by the background of the band-to-band transitions introduced in **Fig. 1**. In order to derive this particular and most suited type of transients, we first recorded the temporal decays at the spectral positions of $X^L$ and $X^U$ yielding an overlap between the temporal intensity evolution of the background and the collexon complex of choice, requiring a correction in order to derive the unperturbed decay-times of $X^L$ and $X^U$. Second, transients that only include the background luminescence were recorded at spectrally most suited positions. As the temporal behaviour of the background luminescence scales with the emission energy (shortens with emission energy, i.e. towards the Fermi edge of the band-to-band-transitions) we chose to record the exclusive decay of the band-to-band transitions at an energy spacing of twice the full width at half maximum of the individual $X^L$ and $X^U$ emission peak. Consequently, we subtracted both sets of transients in order to extract the differential transients of $X^L$ and $X^U$.

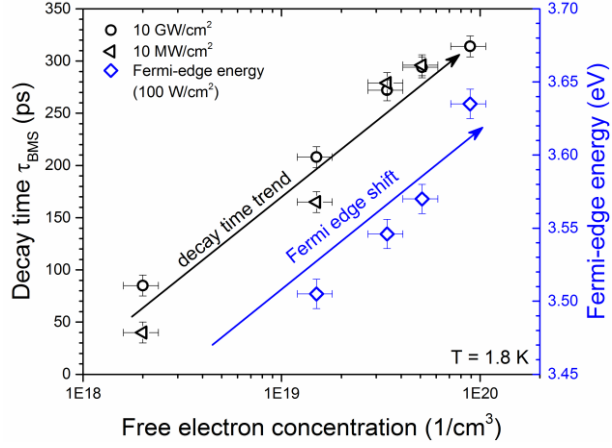

**Fig. S7 | Fermi edge analysis.** The decay-times $\tau_{BMS}$ (black symbols) at the Fermi-edge of the band-to-band transitions (Burstein-Moss-shifted contribution) increases with rising doping concentration. Hence, even at the highest doping concentration we do not observe a strong contribution of compensation mechanisms that would otherwise inverse this trend as commonly observed for other semiconductor compounds.[35] At low doping concentrations, we can still observe a variation of the decay-times depending on the excitation power depending on the ratio between optically excited charge carriers and free electrons introduced by the doping. Naturally, towards higher doping concentrations this decay-time variation vanishes. The energetic position of the Fermi-edge (blue symbols) reflects the free electron concentration introduced by the Ge-doping.

Please note that it is a clear prerequisite of this subtraction to record transients over several orders of magnitude (five to six in our case) in order to achieve a sufficient signal/noise-ratio, cf. **Fig. 2a**. The temporal evolution of the high energy luminescence affected by the Burstein-Moss-shift (BMS) can most efficiently be compared between the analysed samples if the Fermi-edge is chosen as the spectral position. **Fig. S7** depicts such a comparison of $\tau_{BMS}$ for two excitation powers (10 MW/cm$^2$ and 10 GW/cm$^2$) along with the spectral position of the Fermi edge determined under low-excitation power conditions (100 W/cm$^2$ - identical value as in **Fig. 1**). The latter choice is of special importance as optically excited charge carriers also shift the high-energy edge to a certain extend depending on the doping concentration. With rising free electron concentration $\tau_{BMS}$ rises as a clear sign of negligible compensation, cf. **Supplementary Note 2**. Pauli-blocking occurs in the conduction band with rising free electron concentration, leading to a corresponding k-space filling that prolongs the measured decay-times. This is a significant observation as an increasing doping concentration is typically accompanied by diminishing decay-times.[35]

**4. Photoluminescence excitation spectroscopy**
In **Fig. 2b** we show a shift in between the maxima of the PLE spectra related to $X^U$ and the corresponding PL features, which rises towards higher energies with increasing free electron concentration. In addition, a similar, electron-density-dependent shift can be observed for all excitation channels related to $X_B^U$, $X_{n=2}^U$ and $X_C^U$. Interestingly, the PLE maximum related to $X^L$ is also shifted towards higher energies when compared to its PL counterpart as noticeable in the PLE spectrum associated to the defect band situated at 3.425 eV. Similar shifts between PLE and PL maxima have been attributed to self-absorption processes often related to donor-acceptor-pair luminescence in, e.g., GaN[27] and CdS.[36] In our case, any excitation energy in close resonance to $X^U$ promotes the generation of $X^L$ related complexes scaling with free electron concentration due to an intensification of the closely related scattering process quantified by means of $\tau_R^L$, cf. **Fig. 2a** and **3**. However, the pronounced generation of $X^U$ complexes at the GaN:Ge sample's surface also promotes the absorption of light that arises from the decay of $X^L$ by means of $X^U$ dissociation - a process similar to infrared absorption of free excitons in Si and Ge that shows an absorption extension towards the free excitonic continuum.[37,38] Hence, the most efficient generation of $X^L$ via the total resonant excitation of $X^U$ complexes is not necessarily accompanied by a maximum in PL intensity as the entire process is counterbalanced by a self-absorption process. Finally, it is exactly this subtle balance between close-to resonance excitation and self-absorption processes that governs the particular structure of the PLE spectra shown in **Fig. 2**. In this picture it is also possible to understand the electron-density-dependency of the maximum in the PLE spectra related to $X^L$. As long as the scattering process that feeds $X^L$ via $X^U$ is of limited efficiency due to a comparably low number of free electrons (3.4E19/cm$^3$), the self-absorption process is negligible and only a minor shift in between the maximum in the PLE and PL spectrum occurs. However, as soon as the free electron concentration rises, the resonant excitation of $X^L$ via $X^U$ becomes increasingly efficient as the underlying scattering processes are enhanced (5.1E19/cm$^3$ and 8.9E19/cm$^3$). A closer look to the PLE spectrum of the defect-related band (3.425 eV) in **Fig. 2a** even shows that the offset in between the PLE and PL maxima is here more pronounced for $X^L$. Hence, $X^L$ constitutes the supreme excitation channel of this defect-band as the strongest sign of a self-absorption process is observed, despite the fact that both excitation channels related $X^U$ and $X^L$ show similar intensity in the PLE spectrum. A detailed analysis of the microscopic origin that balances the intensity ratio for this defect-band goes beyond the scope of the present manuscript and must remain a task for future work.

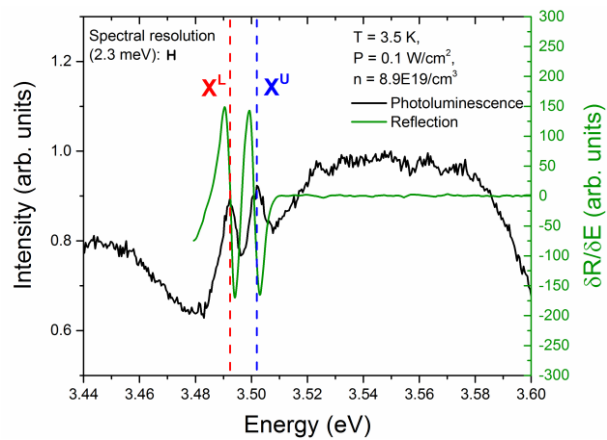

**Fig. S8 | Luminescence vs. reflectivity.** The occurrence of $X^U$ and $X^L$ in the photoluminescence spectrum (black) is directly accompanied by a pronounced change in reflection (normalised). This observation does not only exclude that $X^U$ and $X^L$ are related to any Ge-bound excitonic luminescence but also precludes an effect of the GaN buffer layer. Only complexes similar to free excitons evoke such strong reflection changes - in our case represented by a collexon and bicollexon.

**Fig. S8** shows a data compilation that was extracted from our polychromatic PLE data that we not only recorded with a dye-laser but also with a conventional XBO-lamp. The figure contains an extracted PL spectrum from above Fermi-edge excitation (black line) and reflectivity data (green line). As the back-reflected lamp intensity is continuously monitored during the recording of the polychromatic PLE spectra, a reflectivity spectrum can straightforwardly be extracted. Here, a pronounced change in reflectivity is observed in close energetic vicinity to the spectral position of $X^U$ and $X^L$ in the PL spectrum, cf. **Fig. S8**. Such behaviour is known for complexes like free excitons causing major reflectivity changes in contrast to classical bound excitons.[39] Hence, our observation from **Fig. S8** supports our interpretation of $X^U$ and $X^L$ as a novel type of quasiparticles unaffected by impurities, like, e.g., Ge-donors. Furthermore, **Fig. S8** indirectly proves that neither $X^L$ nor $X^U$ arise from the buffer layer underneath the GaN:Ge layer as discussed in **Supplementary Note 1**. Any luminescence contribution of the GaN buffer layer would always be dominated by bound-excitonic luminescence that stands in total contrast to the strong change in reflectivity directly observed in **Fig. S8**. At this point, low-temperature ellipsometry measurements represents a task for future work.

**5. Temperature-dependent Photoluminescence**

The exciton-like character of $X^L$ and $X^U$ is confirmed by their rapid thermalization behaviour as shown in the inset of **Fig. S9**. Both emission bands vanish simultaneously, while the Burstein-Moss-shifted background luminescence of band-to-band transitions remains comparably stable. Here, **Fig. S9** summarizes the corresponding thermalization behaviour of $X^L$, $X^U$, and BMS in an Arrhenius-plot. The trend of diminishing intensity with rising temperature is fitted in accordance to Ref. 40 under consideration of two thermal activation energies. The background of the band-to-band transitions is most robust against all thermalization processes, while $X^U$ and $X^L$ thermalize more rapidly. This behaviour can be understood as soon as the particular physical origin of $X^L$ and $X^U$ as a bicollexon and a collexon is taken into account. The liberation of one of the constituents (electron or hole) of the individual many-particle complex always leads to a complete dissociation of $X^L$ or $X^U$. Hence, an increasing number of particles that form the underlying excitonic complex (collexon → bicollexon) increases the susceptibility to thermalization. Hence, $X^L$ exhibits smaller thermal activation energies than $X^U$, while the BMS luminescence constitutes the most simplistic and therefore also most resistant luminescence. More detailed studies of this particular temperature dependency must be undertaken with focus on the doping and excitation power - another motivation for future research. A high temperature PL spectrum (100 K) of the collexons is shown in the inset of **Fig. S1**.

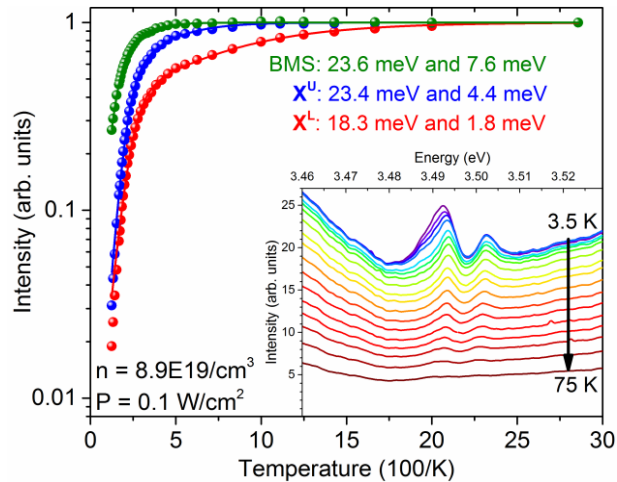

**Fig. S 9 | Temperature-dependent luminescence.** The luminescence associated to $X^L$ and $X^U$ shows a complex temperature dependency (inset) before both emission lines vanish simultaneously at a temperature of 75 K. The detailed analysis of the thermalization behaviour of $X^L$, $X^U$, and the Burstein-Moss-shifted (BMS) background luminescence, reveals different activation energies. Here, BMS is most robust against thermalization mechanisms, while $X^U$ and $X^L$ thermalize more rapidly depending on the number of particles forming the underlying excitonic complex.

## SUPPLEMENTARY REFERENCES

1. Bechstedt, F. *Many-Body Approach to Electronic Excitations*. (Springer, 2015).

2. Hedin, L. New method for calculating the one-particle Green's function with application to the electron-gas-problem. *Phys. Rev.* **139,** A796 (1965).

3. Shklovskii, B. I. & Efros, A. L. Electronic Properties of Doped Semiconductors. *Springer Ser. Solid-State Sci.* **45,** 72–93 (1984).

4. Rohlfing, M. & Louie, S. Electron-hole excitations and optical spectra from first principles. *Phys. Rev. B* **62,** 4927–4944 (2000).

5. Schleife, A., Rödl, C., Fuchs, F., Hannewald, K. & Bechstedt, F. Optical absorption in degenerately doped semiconductors: Mott transition or Mahan excitons? *Phys. Rev. Lett.* **107,** 236405 (2011).

6. Finkelstein, G., Shtrikman, H. & Bar-Joseph, I. Optical Spectroscopy of a Two-Dimensional Electron Gas near the Metal-Insulator Transition. *Physical Review Letters* **74,** 976–979 (1995).

7. Berggren, K. F. & Sernelius, B. E. Band-gap narrowing in heavily doped many-valley semiconductors. *Phys. Rev. B* **24,** 1971–1986 (1981).

8. Feneberg, M. *et al.* Band gap renormalization and Burstein-Moss effect in silicon- and germanium-doped wurtzite GaN up to 1020 cm-3. *Phys. Rev. B - Condens. Matter Mater. Phys.* **90,** 1–10 (2014).

9. Mott, N. F. Metal-insulator transition. *Rev. Mod. Phys.* **40,** 677–683 (1968).

10. Mahan, G. D. Excitons in degenerate semiconductors. *Phys. Rev.* **153,** 882–889 (1967).

11. Esser, A., Zimmermann, R. & Runge, E. Theory of trion spectra in semiconductor nanostructures. *Phys. Status Solidi Basic Res.* **227,** 317–330 (2001).

12. Suris, R. A. in *Optical Properties of 2D Systems with Interacting Electrons SE - 9* (eds. Ossau, W. & Suris, R.) **119,** 111–124 (Springer Netherlands, 2003).

13. Haug, Hartmut (Goethe-Universität Frankfurt, Germany), Koch, Stephan W (Philipps-Universität Marburg, G. *Quantum Theory of the Optical and Electronic Properties of Semiconductors*. (World Scientific, 2009).

14. de Carvalho, L. C., Schleife, A. & Bechstedt, F. Influence of exchange and correlation on structural and electronic properties of AlN, GaN, and InN polytypes. *Phys. Rev. B* **84,** 1–13 (2011).

15. Furthmüller, J. *et al.* Band structures and optical spectra of InN polymorphs: Influence of quasiparticle and excitonic effects. *Phys. Rev. B* **72,** 205106 (2005).

16. Brunhes, T., André, R., Arnoult, a., Cibert, J. & Wasiela, a. Oscillator strength transfer from X to X+ in a CdTe quantum-well microcavity. *Phys. Rev. B* **60,** 11568–11571 (1999).

17. Rapaport, R., Qarry, a., Cohen, E., Ron, a. & Pfeiffer, L. N. Charged excitons and cavity polaritons. *Phys. Status Solidi Basic Res.* **227,** 419–427 (2001).

18. Suris, R. A. Correlation Between Trion and Hole in Fermi Distribution in Process of Trion Photo-Excitation in Doped QWs. *Opt. Prop. 2D Syst. with Interact. Electrons SE - 9* **119,** 111–124 (2003).

19. Suris, R. A. *et al.* Excitons and trions modified by interaction with a two-dimensional electron gas. *Phys. Status Solidi Basic Res.* **227,** 343–352 (2001).

20. Monemar, B. *et al.* Recombination of free and bound excitons in GaN. *Phys. Status Solidi B-Basic Solid State Phys.* **245,** 1723–1740 (2008).

21. Kindel, C. *et al.* Spectral diffusion in nitride quantum dots: Emission energy dependent linewidths broadening via giant built-in dipole moments. *Phys. Status Solidi - Rapid Res. Lett.* **8,** 408–413 (2014).

22. Callsen, G. & Pahn, G. M. O. Identifying multi-excitons in quantum dots: The subtle connection between electric dipole moments and emission linewidths. *Phys. Status Solidi - Rapid Res. Lett.* **9,** 521–525 (2015).

23. Nenstiel, C. *et al.* Germanium - The superior dopant in n-type GaN. *Phys. Status Solidi - Rapid Res. Lett.* **9,** 716–721 (2015).

24. Bertram, F. *et al.* Strain relaxation and strong impurity incorporation in epitaxial laterally overgrown GaN: Direct imaging of different growth domains by cathodoluminescence microscopy and micro-Raman spectroscopy. *Appl. Phys. Lett.* **74,** 359–361 (1999).

25. Christen, J. & Riemann, T. Optical micro-characterization of complex GaN structures. *Phys. Status Solidi Basic Res.* **228,** 419–424 (2001).

26. Muth, J. F. *et al.* Absorption coefficient, energy gap, exciton binding energy, and recombination lifetime of GaN obtained from transmission measurements. *Appl. Phys. Lett.* **71,** 2572 (1997).

27. Callsen, G. *et al.* Optical signature of Mg-doped GaN: Transfer processes. *Phys. Rev. B* **86,** (2012).

28. Özgür, Ü. *et al.* A comprehensive review of ZnO materials and devices. *J. Appl. Phys.* **98,** 1–103 (2005).

29. Reshchikov, M. A. & Morkoç, H. Luminescence properties of defects in GaN. *J. Appl. Phys.* **97,** 61301 (2005).

30. Polian, A., Grimsditch, M. & Grzegory, I. Elastic constants of gallium nitride. *J. Appl. Phys.* **79,** 3343–3344 (1996).

31. Callsen, G. *et al.* Phonon deformation potentials in wurtzite GaN and ZnO determined by uniaxial pressure dependent Raman measurements. *Appl. Phys. Lett.* **98,** (2011).

32. Alemu, A., Gil, B., Julier, M. & Nakamura, S. Optical properties of wurtzite GaN epilayers grown on A-plane sapphire. *Phys. Rev. B* **57,** 3761–3764 (1998).

33. Wagner, M. R. *et al.* Bound excitons in ZnO: Structural defect complexes versus shallow impurity centers. *Phys. Rev. B - Condens. Matter Mater. Phys.* **84,** (2011).

34. Wagner, M. R. *et al.* Effects of strain on the valence band structure and exciton-polariton energies in ZnO. *Phys. Rev. B* **88,** 235210 (2013).

35. Ch. Fricke, R. Heitz, A. Hoffmann, I. B. Recombination mechanisms in highly doped CdS:In. *Phys. Rev. B* **49,** 5313 (1994).

36. Broser, I., Gutowski, J. & Riedel, R. Excitation spectroscopy of the donor-acceptor-pair luminescence in CdS. *Solid State Commun.* **49,** 445–449 (1984).

37. Timusk, T. Far-infrared absorption study of exciton ionization in germanium. *Phys. Rev. B* **13,** 3511 (1976).

38. Timusk, T., Navarro, H., Lipari, N. O. & Altarelli, M. Far-infrared absorption by excitons in silicon. *Solid State Commun.* **25,** 217–219 (1978).

39. Volm, D. *et al.* Exciton fine structure in undoped GaN epitaxial films. *Phys. Rev. B* **53,** 16543–16550 (1996).

40. Bimberg, D., Sondergeld, M. & Grobe, E. Thermal dissociation of excitons bounds to neutral acceptors in high-purity GaAs. *Phys. Rev. B* **4,** 3451 (1971).